\documentclass[aps,prb,showpacs,eqsecnum]{revtex4}
\usepackage{psfig,graphicx}

\newcommand{\diag}{{\rm diag\,}}
\newcommand{\qdet}{{\rm qdet\,}}
\newcommand{\trg}{{\rm trg\,}}
\newcommand{\tr}{{\rm tr\,}}
\newcommand{\detg}{{\rm detg\,}}
\newcommand{\sign}{{\rm sign\,}}

\newcommand{\im}{{\rm Im\,}}

\newcommand{\half}{\frac{1}{2}}

\begin{document}

\title{The $k$--Point Random Matrix Kernels Obtained
       from One--Point Supermatrix Models}
\author{Johan Gr\"onqvist$\dag$ \ , Thomas Guhr$\dag$ \
        and Heiner Kohler$\ddag$}
\affiliation{$\dag$\
         Matematisk Fysik, LTH, Lunds Universitet,
         Box 118, 22100 Lund, Sweden\\
         $\ddag$
         Departamento de Teor\'{i}a de Materia Condensada,
         Universidad Aut\'onoma, Madrid, Spain}
\date{\today}
\begin{abstract}
  The $k$--point correlation functions of the Gaussian Random Matrix
  Ensembles are certain determinants of functions which depend on only
  two arguments. They are referred to as kernels, since they are the
  building blocks of all correlations.  We show that the kernels are
  obtained, for arbitrary level number, directly from supermatrix
  models for one--point functions. More precisely, the generating
  functions of the one--point functions are equivalent to the kernels.
  This is surprising, because it implies that already the one--point
  generating function holds essential information about the $k$--point
  correlations. This also establishes a link to the averaged ratios of
  spectral determinants, i.e.~of characteristic polynomials.
\end{abstract}



\maketitle

\section{Introduction}
\label{sec1}

Random Matrix Theory (RMT) allows one to model a rich variety of
complex systems~\cite{Mehta,Haake,GMGW}. The Gaussian Unitary Ensemble
(GUE) of random matrices is used in the absence of time reversal
invariance. The Gaussian Orthogonal Ensemble (GOE) and the Gaussian
Symplectic Ensemble (GSE) apply if time reversal invariance holds and
if the levels are not or are Kramers degenerate, respectively. These
three cases GOE, GUE and GSE are labeled by the Dyson index
$\beta=1,2,4$.  Supersymmetry~\cite{efe83,VWZ} often yields a
considerably clearer insight into the structure of random matrix
models. This is so, because supersymmetry drastically reduces the
number of degrees of freedom without giving away any information
contained in the model.  Thus, supersymmetry removes a certain kind of
redundancy. Loosely speaking, the supersymmetric formulation plays for
the random matrix model the role of an ``irreducible representation''.

In this contribution, we present an unexpected direct connection
between the $k$--point correlation functions of the three Gaussian
Ensembles and the generating functions of the one--point functions. The
$k$--point correlation functions are determinants (GUE) or quaternion
determinants (GOE,GSE)~\cite{Mehta}. All entries of these determinants
are fully specified by one function of two energy arguments.  Because
of their fundamental importance, these functions are referred to as
kernels. In the supersymmetric formulation, generating functions are
used which, upon derivative with respect to source variables, yield the
$k$--point correlation functions. Especially the generating function of
the one--point correlation function, or rather one--point function,
depends on an energy and on a somewhat unphysical source variable. The
source variable is needed to break a symmetry in the supersymmetric
matrix model.  Hence, the number of dependend variables is the same for
the generating function of the one--point function and for the kernels.
We show in the sequel that, surprisingly, the generating function of
the one--point function is fully equivalent to the kernels. This is
true for all three Gaussian Ensembles and for arbitrary level numbers.
Thus, a fundamental link is established between the one--point
functions and the $k$--point correlation functions.

The article is organized as follows. For the convenience of the reader,
we briefly compile the relevant formulae for the random matrix
correlation functions and kernels in Section~\ref{sec2}. We follow
closely Mehta's book~\cite{Mehta}.  In Section~\ref{sec3}, we present
our main results and discuss implications. The derivations are
performed in Section~\ref{sec4}. Summary and conclusion are given in
Section~\ref{sec5}.

\section{Random Matrix Correlation Functions and Kernels}
\label{sec2}

The $k$--point correlation functions $R_k(x_1,\ldots,x_k)$ are the
probability densities to find $k$ energies at positions
$x_1,\ldots,x_k$, regardless of labeling. They can be written as
averages over a probability density $P_N(H)$ of a $N\times N$ Hamilton
matrix $H$,
\begin{equation}
R_k(x_1,\ldots,x_k) = \int P_N(H) \prod_{p=1}^k
                          \tr\delta(x_p-H) d[H] \ ,
\label{eq2.0}
\end{equation}
where $d[H]$ is the volume element, that is, the product of the
differentials of all independent variables in $H$. We mention in
passing that this definition contains contributions proportional to
$\delta(x_p-x_q)$. However, as this issue is not important here, we
ignore it and refer the reader to the discussion of those details in
Ref.~\cite{GMGW}.

For the Gaussian Ensembles, the correlation functions have a remarkable
determinant structure~\cite{Mehta}. All knowledge needed to construct
the full function $R_k^{(\beta)}(x_1,\ldots,x_k)$ is contained in one
single function, the kernel, which depends on two energy arguments. In
the case of the GUE ($\beta=2$) for $N$ levels one has
\begin{equation}
R_k^{(2)}(x_1,\ldots,x_k) =
       \det\left[K_N^{(2)}(x_p,x_q)\right]_{p,q=1,\ldots,k} \ ,
\label{eq2.1}
\end{equation}
where the kernel is given by
\begin{equation}
K_N^{(2)}(x_p,x_q) = \sum_{n=0}^{N-1} \varphi_n(x_p)\varphi_n(x_q) \ .
\label{eq2.2}
\end{equation}
Here, $\varphi_n(z)$ denotes the oscillator wave function
\begin{equation}
\varphi_n(z) = \frac{1}{\sqrt{2^nn!\sqrt{\pi}}}
               \exp\left(-\frac{z^2}{2}\right) H_n(z)
\label{eq2.3}
\end{equation}
and $H_n(z)$ is the Hermite polynomial of order $n$~\cite{Abram}.

Due to the additional symmetries, the corresponding expressions for the
GOE ($\beta=1$) and the GSE ($\beta=4$) are more involved.  For the
GOE, the kernel is given by
\begin{equation}
K_N^{(1)}(x_p,x_q) = K_N^{(2)}(x_p,x_q)
                    \quad  + \sqrt{\frac{N}{2}}
                 \varphi_{N-1}(x_p)\int_{-\infty}^{+\infty}
                 \varepsilon(x_q-z)\varphi_N(z)dz + \alpha_N(x_p) \ ,
\label{eq2.4}
\end{equation}
where $K_N^{(2)}(x_p,x_q)$ is the GUE kernel and the function
\begin{equation}
\alpha_N(x)=\left\{\begin{array}{cc}
  \varphi_{N-1}(x)/\int_{-\infty}^{+\infty}\varphi_{N-1}(t)dt
  & \qquad N\ {\rm odd}\cr
  0&\qquad N\ {\rm even}\end{array}\right.
\label{eq2.51}
\end{equation}
enters.  We also use the notation
\begin{equation}
\varepsilon(z) = \frac{1}{2} \sign(z) \ . \label{eq2.5}
\end{equation}
In the case of the GSE, the kernel reads
\begin{equation}
K_N^{(4)}(x_p,x_q)=\frac{1}{\sqrt{2}}
                 K_{2N+1}^{(2)}(\sqrt{2}x_p,\sqrt{2}x_q)
                                   \nonumber\\
                  + \sqrt{\frac{2N+1}{2}}
                 \varphi_{2N}(\sqrt{2}x_p)\int_{-\infty}^{+\infty}
                 \varepsilon(x_q-z)\varphi_{2N+1}(\sqrt{2}z)dz \ ,
\label{eq2.65}
\end{equation}
the first two terms are the same as in the GOE kernel, but for $2N+1$
levels. However, the function $\alpha_N(x)$ does not appear. It is
convenient to scale the energy arguments with $\sqrt{2}$.

The ordinary determinant for the GUE correlation functions is replaced
by quaternion determinants. One has for the GOE
\begin{eqnarray}
R_k^{(1)}(x_1,\ldots,x_k) =
            \qdet\left[ \begin{array}{cc}
            K_N^{(1)}(x_p,x_q) & DK_N^{(1)}(x_p,x_q)\cr
            JK_N^{(1)}(x_p,x_q) &  K_N^{(1)}(x_q,x_p)
            \end{array} \right]_{p,q=1,\ldots,k}
\label{eq2.6}
\end{eqnarray}
and, similarly, for the GSE
\begin{eqnarray}
R_k^{(4)}(x_1,\ldots,x_k) =
            \qdet\left[ \begin{array}{cc}
            K_N^{(4)}(x_p,x_q) & DK_N^{(4)}(x_p,x_q)\cr
            IK_N^{(4)}(x_p,x_q) &  K_N^{(4)}(x_q,x_p)
            \end{array} \right]_{p,q=1,\ldots,k} \ .
\label{eq2.7}
\end{eqnarray}
Here, $D,I$ and $J$ are certain derivative and integral operators,
respectively. We write $K_N^{(2)}(x_p,x_q)$ for the GUE kernel which
Mehta denotes by $K_N(x_p,x_q)$~\cite{Mehta}. Mehta works with the
kernel $S_N(x_p,x_q)$ for the GOE and the GSE. This function
$S_N(x_p,x_q)$ is our $K_N^{(1)}(x_p,x_q)$ without the function
$\alpha_N(x_p)$.  We decided to introduce the kernels
$K_N^{(1)}(x_p,x_q)$ and $K_N^{(4)}(x_p,x_q)$, because the function
$\alpha_N(x_p)$ is only present in the GOE, but not in the GSE case. As
we will show, the kernels $K_N^{(\beta)}(x_p,x_q)$ are the ones that
appear naturally in the supersymmetry context. More information on the
relation between Mehta's kernels and the kernels
$K_N^{(\beta)}(x_p,x_q)$ and on how they enter the
expressions~(\ref{eq2.6}) and~(\ref{eq2.7}) for the correlation
functions can be found in \ref{app0}.

In concluding this compilation, we underline once more that the
knowledge of the three kernels suffices to build up all $k$--point
correlation functions for the three ensembles GOE, GUE and GSE.

\section{Kernels, Matrix Integrals, Generating Functions
         and Random Matrix Averages}
\label{sec3}

Surprisingly, one can obtain the kernels from the lowest dimensional
one--point supermatrix models which reflect the appropriate symmetries.
This and its implications state the main result of the present
contribution.  For the GUE, i.e.~for $\beta=2$, we have
\begin{equation}
K_N^{(2)}(x_q,x_p) =
             \frac{1}{\pi}\frac{\exp\left(x_p^2/2-x_q^2/2\right)}
                      {x_p-x_q}\
              \im\left(\frac{1}{2}\int \exp\left(-\trg\sigma^2\right)
                      \detg^{-N}(\sigma-x^-) d[\sigma]
                 -1 \right) \ ,
\label{eq3.1}
\end{equation}
where
\begin{equation}
\sigma=\left[\begin{array}{cc}a&\lambda^*\\
                     \lambda&ib\end{array}\right]
\label{herm}
\end{equation}
is a $2\times 2$ Hermitean supermatrix. The entries $a,b$ are real
commuting and $\lambda$ is complex anticommuting.  The energies are
ordered in the diagonal matrix $x=\diag(x_p,x_q)$. The variable $x_p$
is supplemented with a small imaginary increment $i\eta$ such that
$x_p^-=x_p-i\eta$ and $x^-=\diag(x_p^-,x_q)$.  The corresponding result
in the case of the GOE, i.e.~for $\beta=1$, reads, for even and odd
level number $N$,
\begin{equation}
K_N^{(1)}(x_q,x_p) =
        \frac{1}{\pi}\frac{\exp\left(x_p^2/2-x_q^2/2\right)}
                      {x_p-x_q}\
           \im\left(\frac{1}{8}\int
           \exp\left(-\frac{1}{2}\trg\sigma^2\right)
                      \detg^{-N/2}(\sigma-x^-) d[\sigma]
                 -1 \right) \ .
\label{eq3.2}
\end{equation}
Finally, for the GSE, i.e.~for $\beta=4$, we have
\begin{equation}
K_N^{(4)}(x_q,x_p) =
             \frac{1}{2\pi}\frac{\exp\left(x_q^2-x_p^2\right)}
                           {x_q-x_p}\
        \im\left(\frac{1}{8}\int \exp\left(-\trg\sigma^2\right)
                      \detg^{-N}(\sigma-x^-) d[\sigma]
                 -1 \right) \ .
\label{eq3.3}
\end{equation}
In the cases of the GOE and the GSE, $\sigma$ is a $4\times 4$
Hermitian supermatrix with an additional symmetry, often referred to as
orthosymplectic \cite{efe83,VWZ}. Explicitly, $\sigma$ reads
\begin{equation}
\sigma=\left[\begin{array}{cccc}\sqrt{c}a&\sqrt{c}d&\lambda^*&-\lambda\\
                     \sqrt{c}d&\sqrt{c}b&\mu^*&-\mu\\
                     \lambda&\mu&\sqrt{-c}w&0\\
                     \lambda^*&\mu^*&0&\sqrt{-c}w\end{array}\right]\quad ,
\label{orthosymm}
\end{equation}
where $c=1$ or $c=-1$ for GOE and GSE, respectively. Here the variables
$a,b,d$ and $w$ are real commuting, while $\lambda,\lambda^*$ and
$\mu,\mu^*$ are complex anticommuting. The diagonal matrix of the
energy arguments now also has dimension $4\times 4$ and reads
$x=\diag(x_p,x_p,x_q,x_q)$. However, for brevity we always write $x$
for the $2\times 2$ and for the $4\times 4$ energy matrix. All formulae
given here are exact for finite values of $N$. Thus, as always,
supersymmetry decouples the number of integrations to be done from the
level number $N$. The number of integrations is fixed, while the level
number $N$ may take arbitrary values.  The common form of the three
results~(\ref{eq3.1}) to~(\ref{eq3.3}) is evident. The differences for
the three Gaussian ensembles lie in the structure of the matrices
$\sigma$. Our definitions and notations are the standard
ones~\cite{efe83,VWZ} and, in particular, they follow the definitions
and notations of Refs.~\cite{GKP2,GK2,Gu1}. Thus, we may even formulate
the results~(\ref{eq3.1}) to~(\ref{eq3.3}) in the compact form
\begin{equation}
K_N^{(\beta)}(x_q,x_p) =
             \frac{1}{\gamma\pi}
             \frac{\exp\left(\gamma
                       \left(x_p^2-x_q^2\right)/2\right)}
                       {x_p-x_q}\ \im\left(\frac{\beta^2}{8\gamma^4}
           \int \exp\left(-\frac{\beta}{2\gamma}\trg\sigma^2\right)
                      \detg^{-\beta N/2|\gamma|}(\sigma-x^-) d[\sigma]
                 -1 \right) \ ,
\label{eq3.3a}
\end{equation}
where we introduced $\gamma=1$ for $\beta=1,2$ and $\gamma=-2$ for
$\beta=4$.

In spite of its non--trivial character, the result~(\ref{eq3.1}) is
easily proven because it is an immediate consequence of an integral
representation of the kernel $K_N^{(2)}(x_p,x_q)$ which was found in
Ref.~\cite{Gu1}. We briefly sketch the derivation in
Section~\ref{sec4}. In fact, expressions similar to Eq.~(\ref{eq3.1})
have already been used for a study involving the chiral GUE~\cite{GW}
and for a certain generalization of the GUE~\cite{GS}. On the other
hand, the proofs of the results~(\ref{eq3.2}) and~(\ref{eq3.3}) are
more involved. They will also be given in Section~\ref{sec4}.

Formulae~(\ref{eq3.1}) to~(\ref{eq3.3}) are remarkable, because they
establish a direct and previously unknown connection between the
kernels and the generating functions $Z_1^{(\beta)}(\bar{x})$ of the
one--point functions,
\begin{equation}
\widehat{R}_1^{(\beta)}(x_1) = \frac{1}{2|\gamma|\pi}
                               \frac{\partial}{\partial J_1}
                               Z_1^{(\beta)}(\bar{x})\Bigg|_{J_1=0} \ .
\label{eq3.4}
\end{equation}
We introduced the diagonal matrix $\bar{x}=\diag(x_1-J_1,x_1+J_1)$
which contains the energy argument $x_1$ and the source variable $J_1$.
To be consistent with the previous notation, we use $x_1$ to denote the
argument of the one--point functions. The one--point function is
written as $\widehat{R}_1^{(\beta)}(x_1)=
\widetilde{R}^{(\beta)}_1(x_1)+iR^{(\beta)}_1(x_1)$ such that the level
density is the imaginary part,
$\im\widehat{R}_1^{(\beta)}(x_1)=R^{(\beta)}_1(x_1)$. As is well known,
the matrix integrals in the expressions~(\ref{eq3.1}) to~(\ref{eq3.3})
are precisely the generating functions,
\begin{equation}
Z_1^{(\beta)}(x) = \frac{\beta^2}{8\gamma^4}
                     \int \exp\left(-\frac{\beta}{2|\gamma|}
                             \trg\sigma^2\right)
                   \detg^{-\beta N/2|\gamma|}(\sigma-x^-) d[\sigma]
\label{eq3.5}
\end{equation}
with the appropriate supermatrices $\sigma$.  Thus, we arrive at
\begin{eqnarray}
K_N^{(2)}(x_q,x_p) &=&\frac{1}{\pi}\exp\left(x_p^2/2-x_q^2/2\right)
                 \im\frac{Z_1^{(2)}(x)-Z_1^{(2)}(0)}{x_p-x_q} \ ,
                              \nonumber\\
K_N^{(1)}(x_q,x_p) &=&\frac{1}{\pi} \exp\left(x_p^2/2-x_q^2/2\right)
                 \im\frac{Z_1^{(1)}(x)-Z_1^{(1)}(0)}{x_p-x_q} \ ,
                              \nonumber\\
K_N^{(4)}(x_q,x_p) &=& \frac{1}{2\pi}\exp\left(x_q^2-x_p^2\right)
                 \im\frac{Z_1^{(4)}(x)-Z_1^{(4)}(0)}
                                   {x_q-x_p} \ .
\label{eq3.6}
\end{eqnarray}
For the GSE kernel $K_N^{(4)}(x_q,x_p)$ the arguments of the
exponential are interchanged with respect to the GUE kernel
$K_N^{(2)}(x_q,x_p)$ and the GOE kernel $K_N^{(1)}(x_q,x_p)$. We notice
that $Z_1^{(\beta)}(x)$ depends on the two energies $x_p$ and $x_q$.
There is no source variable here. Moreover, we have
$Z_1^{(\beta)}(0)=1$ due to the definition of the generating function.
Again, we can write
\begin{eqnarray}
K_N^{(\beta)}(x_q,x_p) =
             \frac{1}{\gamma\pi}
             \exp\left(\frac{\gamma}{2}
                       \left(x_p^2-x_q^2\right)\right)
             \im\frac{Z_1^{(\beta)}(x)-Z_1^{(\beta)}(0)}
                                   {x_q-x_p}
\label{eq3.6a}
\end{eqnarray}
which combines the three results~(\ref{eq3.6}) in a compact form.

Formulae~(\ref{eq3.6}) state a close connection between the kernels and
the generating functions. The kernels can be viewed as difference
quotients of the generating functions at the two points $x$ and $0$.
The crucial quantity is the difference $x_p-x_q$. By construction, the
generating functions are unity whenever the two arguments degenerate.
Thus, $Z_1^{(\beta)}(x)$ moves away from unity as function of
$x_p-x_q$.  If one takes the limit $x_q\to x_p$, the difference
quotient becomes the differential quotient~(\ref{eq3.4}). This yields
the kernels at $x_p=x_q$, that is, the level densities as function of
the single remaining variable. In \ref{app2} we discuss extensions of
the previous results if real parts contribute to the correlation
functions.

To further clarify the meaning of these findings, we rewrite the
generating functions as averages over the original random matrices in
ordinary space,
\begin{equation} Z_1^{(\beta)}(x) = C_{N\beta}
                   \int \exp\left(-\frac{\beta}{2}\tr H^2\right)
   \left(\frac{\det(H-x_q)}{\det(H-x_p^-)}\right)^{|\gamma|} d[H] \ ,
\label{eq3.7}
\end{equation}
with normalization constants $C_{N\beta}$.  Here, the matrices $H$
parameterize the GOE, GUE and GSE for $\beta=1,2,4$, respectively, as
defined in Mehta's book~\cite{Mehta}. Combining Eqs.~(\ref{eq3.6})
and~(\ref{eq3.7}), we see that the kernels themselves are, apart from
factors, averages over the Gaussian ensembles.  This is a surprising
insight. According to the definition~(\ref{eq2.0}), the $k$--point
correlation function of a Gaussian ensemble is one single matrix
integral for a fixed value of $k$. The results presented here imply
that this single average breaks up into products of averages. This is
intimately related to the determinant structure, but it is a stronger
statement because it identifies the determinant structure as stemming
from the break up of the random matrix average.

Furthermore, it follows from Eqs.~(\ref{eq3.6}) and~(\ref{eq3.7}) that
the random matrix kernels are essentially an average over a ratio of
spectral determinants, taken at the two different energies. This
relates our findings to the presently much discussed issue of
characteristic polynomials, spectral determinants and their moments,
see Refs.~\cite{Str,Baik} and references therein. For matrix dimension
$N=2$, the connection between averages over the ratio of two
characteristic polynomials and the kernel was recently observed in
Ref.~\cite{BrHi} in the GOE case ($\beta=1$).

\section{Kernels Expressed as Eigenvalue Integrals
         in Superspaces}
\label{sec4}

We prove the results in the previous Section by explicit calculation.
Alternatively, one could try to employ Dyson's Brownian
motion~\cite{Mehta} and its supersymmetric extension~\cite{Gu3} for the
stationary case, i.e.~for the pure ensembles. However, this would still
leave one with the problem of fixing the boundary conditions in an
unambiguous way. Another strategy could consist in showing that the
supermatrix models satisfy the same equations that the kernels obey,
such as the convolution condition. Once more, one is confronted with
some ambiguity. Thus, we believe that the most direct proof is probably
an explicit caculation, but we certainly do not exclude that other
direct proofs also exist.

There are two possibilities to proceed with an explicit calculation.
First, due to the small dimensions of the supermatrices in
Eq.~(\ref{eq4.1}), one can expand the superdeterminants in the
supermatrix models and integrate out the Grassmann variables by ``brute
force".  The resulting expressions are rather complicated and the
calculations to follow are quite cumbersome. Second, one can introduce
eigenvalue--angle coordinates and integrate in a first step over the
supergroups and in a second one over the eigenvalues. We present this
approach in the sequel because the eigenvalue integrals to be solved
here are of a general type which will always appear in exact
calculations involving supersymmetry. In particular, they will show up
in generalizations of the present supermatrix models. Thus, we want to
develop techniques for how to handle them.

We denote the right hand sides of the formulae~(\ref{eq3.1})
to~(\ref{eq3.3}) by
\begin{equation}
L_N^{(\beta)}(x_p,x_q) = \frac{\sqrt{|\gamma|}}{\gamma\pi}
                           \frac{1}{x_p-x_q}\nonumber\\
             \ \im\left(\frac{\beta^2}{8\gamma^4}
\int \exp\left(-\frac{\beta}{2|\gamma|}\trg\sigma^2\right)
                   \detg^{-\beta N/2|\gamma|}(\sigma-x^-) d[\sigma]
                 -1 \right) \ ,
\label{eq4.1}
\end{equation}
where, for notational convenience, the exponential functions are not
included. For the same reason, we also split off a factor
$\sqrt{\gamma}$.  The strategy for the ensuing calculation is now the
same in all three cases $\beta=1,2,4$. The supermatrix integral yields
two contributions, an Efetov--Wegner term~\cite{Roth} and an eigenvalue
integral. The Efetov--Wegner term gives a constant, canceling the unity
to be subtracted in the bracket. To obtain the eigenvalue integral, we
shift the integration matrix $\sigma$ by the diagonal matrix $x$,
introduce eigenvalue--angle coordinates $\sigma=u^{-1}su$ for the
shifted integration matrix and perform the group integral over the
diagonalizing matrix $u$. For $\beta=2$, the group integral is the well
known supersymmetric Itzykson--Zuber integral~\cite{Gu1}. It is trivial
here, because $u$ is only a $2\times 2$ supermatrix.  For $\beta=1,4$
the group integral is over $4\times 4$ supermatrices $u$ and thus
non--trivial. In both cases $\beta=1,4$, it is the same group integral
which can be viewed as a supersymmetric extension of Gelfand's
spherical functions. This integral was first calculated in
Refs.~\cite{GKP2,GK2}. (Responding to some confusion in the literature,
we mention in passing that the integrals involving $8\times 8$
supermatrices were also first solved in these references.  However, in
the present case, the $4\times 4$ case suffices.)  After the group
integrations, we are left with integrals over the eigenvalues
$s=\diag(s_1,is_2)$ for $\beta=2$, $s=\diag(s_{11},s_{21},is_2,is_2)$
for $\beta=1$ and $s=\diag(s_1,s_1,is_{12},is_{22})$ for $\beta=4$.
These eigenvalue integrals are solved in the sequel. We notice that the
eigenvalues in the fermion--fermion block of the supermatrix $\sigma$
carry an imaginary unit. It is due to a Wick rotation which is
necessary to ensure convergence of the supermatrix integrals. Thus,
this imaginary unit has to be ignored when taking the imaginary part on
the right hand side of Eq.~(\ref{eq4.1}).

\subsection{Gaussian Unitary Ensemble}
\label{sec41}

The energy difference $x_p-x_q$ drops out and the matrix integral
reduces to the double integral
\begin{equation}
L_N^{(2)}(x_p,x_q) =
        -\frac{1}{\pi^2}\int_{-\infty}^{+\infty}\int_{-\infty}^{+\infty}
                         \frac{ds_1ds_2}{s_1-is_2}
         \exp\left(-(s_1+x_p)^2+(is_2+x_q)^2\right)\
                         (is_2)^N \im\frac{1}{(s_1^-)^N} \ .
\label{eq4.2}
\end{equation}
This coincides with the double integral found in Ref.~\cite{Gu1}. In
this reference, it was denoted by $C_N(x_p,x_q)$.  However, it is
important to notice that the double integral resulted in
Ref.~\cite{Gu1} from calculating the $k$--point correlation function
for arbitrary $k$, that is, from a group integral over a $2k\times 2k$
unitary supermatrix. In Ref.~\cite{Gu1}, the double integral was
already evaluated and it was shown that
\begin{eqnarray}
K_N^{(2)}(x_p,x_q) = \exp\left(x_p^2/2-x_q^2/2\right)
               L_N^{(2)}(x_p,x_q) \ .
\label{eq4.3}
\end{eqnarray}
This proves formula~(\ref{eq3.1}).

\subsection{Gaussian Orthogonal Ensemble}
\label{sec42}

The orthogonal case has a much more complicated structure. From
Refs.~\cite{GKP2,GK2}, a triple integral results,
\begin{eqnarray}
L_N^{(1)}(x_p,x_q) &=& \frac{1}{8\pi^2}\int_{-\infty}^{+\infty}
                       \int_{-\infty}^{+\infty}
                       \int_{-\infty}^{+\infty}
         \frac{|s_{11}-s_{21}|ds_{11}ds_{21}ds_2}
              {(s_{11}-is_2)^2(s_{21}-is_2)^2}
                       \nonumber\\
                   & & \quad
         \exp\left(-\frac{1}{2}(s_{11}+x_p)^2
                   -\frac{1}{2}(s_{21}+x_p)^2
                    +(is_2+x_q)^2\right)
                       \nonumber\\
                   & & \quad
         \left(2(x_p-x_q)(s_{11}-is_2)(s_{21}-is_2)
               +(s_{11}+s_{21}-2is_2)\right)
                       \nonumber\\
                   & & \quad
         (is_2)^N \im\frac{1}{(s_{11}^-)^{N/2}(s_{21}^-)^{N/2}} \ .
\label{eq4.4}
\end{eqnarray}
Again, the energy difference $x_p-x_q$ in the denominator has been
canceled. The technical difficulty to overcome is twofold. First, the
integration variables are coupled in an inconvenient way, even
involving an absolute value. This happens in the term in front of the
Gaussians which stems from the Jacobian or Berezinian of the
eigenvalue--angle coordinates. Second, the last term contains an
imaginary part of the product of two singularities. It cannot be
interpreted as the product of imaginary parts which would simply yield
derivatives of $\delta$ functions.  Both difficulties can be
circumvented by observing that the GOE kernel contains, according to
Eq.~(\ref{eq2.4}), the GUE kernel. Thus, we split off the latter. To
this end, we reformulate the GUE kernel~(\ref{eq4.2}) as a triple
integral. We write $s_1=s_{11}$ and introduce a dummy integration over
the variable $s_{21}$ by multiplying the integrand with the function
$\delta(s_{11}-s_{21})$.
\begin{eqnarray}
L_N^{(2)}(x_p,x_q) &=&\frac{1}{2\pi^2} \int_{-\infty}^{+\infty}
                       \int_{-\infty}^{+\infty}
                       \int_{-\infty}^{+\infty}
         \frac{\delta(s_{11}-s_{21})
                        ds_{11}ds_{21}ds_2}
              {(s_{11}-is_2)(s_{21}-is_2)}
                       \nonumber\\
                   & & \quad
         \exp\left(-\frac{1}{2}(s_{11}+x_p)^2
                   -\frac{1}{2}(s_{21}+x_p)^2
                    +(is_2+x_q)^2\right)
                       \nonumber\\
                   & & \quad
                      \left(s_{11}+s_{21}-2is_2\right)
         (is_2)^N \im\frac{1}{(s_{11}^-)^{N/2}(s_{21}^-)^{N/2}}
\label{eq4.5}
\end{eqnarray}
In order to subtract this formula from Eq.~(\ref{eq4.4}) we have to do
some integrations by parts in both expressions. More precisely, we use
\begin{equation}
\delta(s_{11}-s_{21})=\frac{1}{2} \left( \frac{\partial}{\partial
  s_{11}} - \frac{\partial}{\partial{s_{21}}} \right)
                            \varepsilon(s_{11}-s_{21})
\label{part1}
\end{equation}
in Eq.~(\ref{eq4.5}). This procedure casts the integrand in
Eq.~(\ref{eq4.5}) into the adequate form to be subtracted from the left
hand side of Eq.~(\ref{eq4.4}).  It is also convenient to do an
integration by parts in Eq.~(\ref{eq4.4}) using
\begin{eqnarray}
&&\left(2\left(x_q-x_p\right)-\frac{\partial}{\partial
is_2}-\frac{\partial}{\partial
            s_{11}}-\frac{\partial}{\partial s_{21}}
            +2is_2-s_{11}-s_{21}\right)\nonumber\\
& &\qquad\qquad \exp\left(-\frac{1}{2}(s_{11}+x_p)^2
   -\frac{1}{2}(s_{21}+x_p)^2 +(is_2+x_q)^2\right)=0\ .
 \label{part2}
\end{eqnarray}
With these adjustments we can subtract $L_N^{(2)}(x_p,x_q)$ from
$L_N^{(1)}(x_p,x_q)$ and obtain
\begin{eqnarray}
M_N^{(1)}(x_p,x_q)&=&L_N^{(1)}(x_p,x_q)-L_N^{(2)}(x_p,x_q)\nonumber\\
                       &=&\frac{N}{8\pi^2}\int_{-\infty}^{+\infty}
                       \int_{-\infty}^{+\infty}
                       |s_{11}-s_{21}|ds_{11}ds_{21}
                       \nonumber\\
                   && \qquad\qquad\quad
                      \exp\left(-\frac{1}{2}(s_{11}+x_p)^2
                       -\frac{1}{2}(s_{21}+x_p)^2\right)\
                   \im\frac{1}{(s_{11}^-)^{N/2+1}(s_{21}^-)^{N/2+1}}
                       \nonumber\\
                   & &\quad\int_{-\infty}^{+\infty}ds_2
                       \exp\left(is_{2}+x_q\right)^2
                       \left(is_2\right)^{N-1}\ .
\label{Omega}
\end{eqnarray}
This expression decouples the three--dimensional integral into a
product of a two--dimensional integral and a one--dimensional integral.
Furthermore, the left hand side of Eq.~(\ref{Omega}) factorizes into a
product of functions each depending only on one energy argument.
Therefore, we can write
\begin{equation}
M_N^{(1)}(x_p,x_q)=\frac{N}{8\pi^2}
                        \omega_N^{(1)}(x_p)\psi_N^{(1)}(x_q)
\label{fac}
\end{equation}
The function $\psi_N^{(1)}$ is simply an integral representation for
the Hermite polynomial. The integration of $\omega_N^{(1)}$ requires
more effort. The relations
\begin{eqnarray}
- \left( \frac{N}{2}
  + 1 \right) \left( \frac{\partial}{\partial x_p} + 2 x_p \right)
  \omega_{N+2}^{(1)}(x_p)&=&
  \frac{\partial}{\partial x_p} \omega_N^{(1)}(x_p) \nonumber \\
  \omega_N^{(1)}(x_p) -\left( \frac{N}{2} + 1 \right)
  \omega_{N+2}^{(1)}(x_p) &=& 4 \pi \frac{(-1)^{N+1}}{(N+1)!}
  H_{N+1}(x_p) \exp (-x_p^2)
\label{rec}
\end{eqnarray}
are used. The second formula above was derived by using
\begin{equation}
  H_N(x_p) = \frac{(-1)^NN!}{\pi}\exp \left(x_p^2\right)
                \im \int_{-\infty}^{\infty}
           \frac{\exp(-(\xi+x)^2)}{(\xi^{-})^{N+1}} d\xi
\label{hermiteRel}
\end{equation}
and by the introduction of a dummy variable similar to
Eq.~(\ref{eq4.5}). By combining these relations, we obtain
\begin{eqnarray}
  \omega_N^{(1)}(x_p) &=& -\exp\left(-x_p^2/2\right)
       \left(\frac{4\pi(-1)^{N}}{N!}\int_{-\infty}^{\infty}
        \varepsilon(x_p - t) H_N(t)
  \exp\left(-\frac{t^2}{2}\right) dt + c_N^{(1)} \right)
                 \nonumber \\
  \psi_N^{(1)}(x_q) &=& \frac{\sqrt{\pi} (-1)^{N-1}}{2^{N-1}}
                                       H_{N-1}(x_q) .
\label{GOEparts}
\end{eqnarray}
The form of the above equations are the expected ones, and the
remaining problem is to calculate the integration constant
$c_{N}^{(1)}$. This tedious calculation is performed in \ref{app1}. We
find
\begin{equation}
c_{N}^{(1)} = \left \{
  \begin{array} {cc} 0 & \mbox{ if $N$ even } \\
    - 4 \pi \: 2^{N/2} / N!! & \mbox{ if $N$ odd }
  \end{array} \right . .
\label{intconst}
\end{equation}
This non--vanishing constant gives rise to a contribution which is
identified with the function $\alpha_N$ defined in Eq.~(\ref{eq2.51}).

Now, Eq.~(\ref{GOEparts}) is rewritten in terms of the oscillator wave
functions $\varphi_n$ defined in (\ref{eq2.3}) and compared to
(\ref{eq2.2}) and (\ref{eq2.4}). We then have
\begin{equation}
  M_N^{(1)}(x_p,x_q) = \exp\left(x_q^2/2-x_p^2/2\right)
                  \left(K_N^{(1)}(x_q,x_p) - K_N^{(2)}(x_q,x_p)\right)
\end{equation}
for all values of $N$. This proves Eq.~(\ref{eq3.2}).

\subsection{Gaussian Symplectic Ensemble}
\label{sec43}

For the GSE the structure of the supermatrix $\sigma$ is almost the
same as for the GOE. The group integral found in Refs.~\cite{GKP2,GK2}
can be applied again. However, boson--boson block and fermion--fermion
block are interchanged with respect to the GOE. As a consequence the
imaginary unit now comes in front of the integration variables
$s_{11},s_{21}$ and the contribution of the superdeterminant in
Eq.~(\ref{eq4.4}) is inverted. With an additional rescaling
$\sigma\rightarrow\sigma/\sqrt{2}$ we obtain
\begin{eqnarray}
L_N^{(4)}(x_p,x_q) &=&\frac{1}{8\pi^2}
                       \int_{-\infty}^{+\infty}
                       \int_{-\infty}^{+\infty}
                       \int_{-\infty}^{+\infty}
         \frac{|s_{11}-s_{21}|ds_{11}ds_{21}ds_2}
              {(is_{11}-s_2)^2(is_{21}-s_2)^2}
                       \nonumber\\
                   & &
         \exp\biggl(\frac{1}{2}\left(is_{11}+\sqrt{2}x_p\right)^2
                   +\frac{1}{2}\left(is_{21}+\sqrt{2}x_p\right)^2
                    -\left(s_2+\sqrt{2}x_q\right)^2\biggr)
                       \nonumber\\
                   & &
         \Bigl(2\left(\sqrt{2}x_q-\sqrt{2}x_p\right)
                (is_{11}-s_2)(is_{21}-s_2)
               +
   \left(is_{11}+is_{21}-2s_2\right)\Bigr)
       \left(is_{11}is_{21}\right)^{N}\im\frac{1}{(s_2^-)^{2N}} \ .
\label{gse1}
\end{eqnarray}
Now we can apply the same method as in the case of the GOE. We arrive
at the decomposition
\begin{eqnarray}
L_N^{(4)}(x_p,x_q)&=&L_{2N}^{(2)}\left(\sqrt{2}x_q,\sqrt{2}x_p\right)
                     +M_N^{(4)}(x_p,x_q)\label{fac12}\\
M_N^{(4)}(x_p,x_q)&=&\frac{2N}{8\pi^2}
                     \omega_N^{(4)}(x_p)\psi_N^{(4)}(x_q) \ .
\label{fac2}
\end{eqnarray}
The functions $\omega_N^{(4)}$ and $\psi_N^{(4)}$ are now given by
\begin{eqnarray}
\psi_N^{(4)}(x_q)&=& \int_{-\infty}^{+\infty}ds_2
                       \exp\left(-\left(s_{2}+\sqrt{2}x_q\right)^2\right)
                       \im\frac{1}{(s_2^-)^{2N+1}}\nonumber\\
\omega_N^{(4)}(x_p)&=&\int_{-\infty}^{+\infty}\int_{-\infty}^{+\infty}
                       ds_{11}ds_{21}|s_{11}-s_{21}|
                   \exp\left(\frac{1}{2}
                        \left(is_{11}+\sqrt{2}x_p\right)^2
            +\frac{1}{2}\left(is_{21}+\sqrt{2}x_p\right)^2\right)
                       \left(is_{11}is_{21}\right)^{N-1}
\label{fac3}
\end{eqnarray}
The function $\psi_N^{(4)}(x_q)$ is easily evaluated by making use of
the identity  Eq.~(\ref{hermiteRel}) for Hermite polynomials
\begin{equation}
\psi_N^{(4)}(x_q)=\frac{\pi}{\left(2N\right)!}
       \exp\left(-2x_q^2\right)H_{2N}\left(\sqrt{2}x_q\right)\ .
\label{symp1}
\end{equation}
The integration of $\omega_N^{(4)}(x_p)$ is a little more tricky. The
procedure follows ideas analogous to those used for the calculation of
$\omega_N^{(1)}(x_p)$ for the GOE in Sec.~(\ref{sec42}) and \ref{app1}.
An integration constant occurs in this case as well. It can be fixed in
the same manner as for the GOE. Here, however, it vanishes for all
values of $N$. One finds
\begin{equation}
\omega_N^{(4)}(x_p) =
-\frac{\sqrt{\pi}}{2^{2N-3}}\exp\left(x_p^2\right)
\int_{-\infty}^{+\infty}\varepsilon\left(\sqrt{2}x_p-t\right)
           \exp\left(-t^2/2\right)H_{2N-1}(t)dt \ .
\label{symp2}
\end{equation}
Inserting these results into Eq.~(\ref{fac12}) and expressing
everything in terms of oscillator wave functions one arrives at
\begin{eqnarray}
L_N^{(4)}(x_p,x_q)&=&\exp\left(x_p^2-x_q^2\right)
         \Biggl(\sum_{n=0}^{2N-1}
       \varphi_n\left(\sqrt{2}x_p\right)\varphi_n\left(\sqrt{2}x_q\right)
          +\Biggr.\nonumber\\
       &&\qquad\Biggl.\sqrt{\frac{2N}{2}}\varphi_{2N}(\sqrt{2}x_q)
       \int_{-\infty}^{+\infty}\varepsilon\left(\sqrt{2}x_p-t\right)
                      \varphi_{2N-1}(t)dt\Biggr)\ .
\label{symp3}
\end{eqnarray}
This is almost the final result. We use the integration formula
\begin{equation}
\sqrt{\frac{2N}{2}} \int_{-\infty}^{+\infty}
            \varepsilon(x-t) \varphi_{2N-1} (t) dt=
 \varphi_{2N}(x) + \sqrt{\frac{2N+1}{2}}
 \int_{-\infty}^{+\infty}\varepsilon(x-t)\varphi_{2N+1}(t)dt
\label{symp4}
\end{equation}
to obtain
\begin{eqnarray}
L_N^{(4)}(x_p,x_q)& = &\exp\left(x_p^2-x_q^2\right)
             \biggl(K^{(2)}_{2N+1}\left(\sqrt{2}x_p,\sqrt{2}x_q\right)
            \biggr.\nonumber\\
 &&\qquad\qquad\biggl.
 +\sqrt{\frac{2N+1}{2}}\varphi_{2N}(\sqrt{2} x_q)
 \int_{-\infty}^{+\infty}\varepsilon\left(\sqrt{2}x_p-t\right)
                               \varphi_{2N+1}(t)
  dt\biggr) \ .
\label{symp5}
\end{eqnarray}
This is exactly our assertion~(\ref{eq3.3}).

\section{Summary and Conclusion}
\label{sec5}

We showed that the generating functions for the one--point functions
yield directly the kernels of the correlation functions in RMT. This is
tantamount to saying that the kernels are given by the lowest
dimensional supermatrix models.  We proved this by explicit
calculations for the Gaussian Ensembles GOE, GUE and GSE. Recent
results for supergroup integrals enter our derivation.  We develop new
techniques for integrals over eigenvalues of supermatrices. This was
another reason for us to prove our results by explicit calculation.

The equivalence between kernels and the generating functions for the
one--point functions is an unexpected, surprising insight.  The
generating functions contain a symmetry breaking, the source term. Our
results demonstrate that this symmetry breaking is intimately related
to the correlations themselves. Hence, the generating functions
comprise much more information than just the one--point functions.
Technically, this becomes apparent in the fact that the source variable
adds, with different signs, to the energy variable. There are
effectively two energy arguments which are then identified with those
of the kernels. The kernels are obtained as the difference quotient of
the generating functions. In the limit of the differential quotient,
one finds the well known relation between one--point functions and
their generating functions.

All our results hold for arbitrary level number. Among other things,
this opens yet another possibility to calculate the kernels in the
limit of large level number on the local scale.  Here, one can do that
by a saddlepoint approximation of the one--point supermatrix models. No
Goldstone modes are present and one finds the kernels for all
correlations from the saddlepoints. As this is a straightforward
exercise, we have not presented it in this contribution.

Our findings are likely to have further extensions. {}From
Ref.~\cite{Gu3}, one easily concludes that the structural relation we
observed carries over, for $\beta=2$, to models in which a fixed matrix
is added to the random matrices. Further investigations are in progress
for the cases $\beta=1$ and $\beta=4$. Our results could have relevance
for field theory as well.

\section*{Acknowledgments}

TG and HK acknowledge financial support from the Swedish Research
Council and from the RNT Network of the European Union with Grant
No.~HPRN--CT--2000-00144, respectively. HK also thanks the division of
Mathematical Physics, LTH, for its hospitality during his visits to
Lund.

\appendix

\section{Mehta's kernels and the
         kernels $K_N^{(\beta)}(x_p,x_q)$}
\label{app0} The fundamental piece in Mehta's notation is the function
\begin{equation}
S_N(x_p,x_q) = K_N^{(2)}(x_p,x_q) + \sqrt{\frac{N}{2}}
                 \varphi_{N-1}(x_p)\int_{-\infty}^{+\infty}
                 \varepsilon(x_q-z)\varphi_N(z)dz\ .
\label{appA1}
\end{equation}
Therefore in his expressions for the GOE correlation functions
$\alpha_N(x)$ appears as an independent quantity
\begin{equation}
R_k^{(1)}(x_1,\ldots,x_k) = \qdet\left[ \begin{array}{cc}
            S_N(x_p,x_q)+\alpha(x_p)& DS_N(x_p,x_q)\cr
            JS_N(x_p,x_q) & S_N(x_q,x_p)+\alpha(x_q)
            \end{array} \right]_{p,q=1,\ldots,k} \ .
\label{AppA2}
\end{equation}
In the GSE result, the function $\alpha(z)$ does not appear,
\begin{equation}
R_k^{(4)}(x_1,\ldots,x_k) = \qdet\left[ \frac{1}{\sqrt{2}}
            \left[\begin{array}{cc}
            S_{2N+1}(\sqrt{2}x_p,\sqrt{2}x_q)  &
                          DS_{2N+1}(\sqrt{2}x_p,\sqrt{2}x_q) \\
            IS_{2N+1}(\sqrt{2}x_p,\sqrt{2}x_q) &
                          S_{2N+1}(\sqrt{2}x_q,\sqrt{2}x_p)
            \end{array}\right] \right]_{p,q=1,\ldots,k} \ .
\label{AppA2a}
\end{equation}
The operators $D,I$ and $J$ are defined as acting on the function
$S(x_p,x_q)$ only,
\begin{eqnarray}
DS_N(x_p,x_q)&=&-\frac{d}{dx_q}S_N(x_p,x_q)\nonumber\\
IS_N(x_p,x_q)&=&\int dt\, \varepsilon(x_p-t)S_N(t,x_q)\nonumber\\
JS_N(x_p,x_q)&=&IS_N(x_p,x_q)+\int^{x_p}_0\alpha(t)dt-\int^{x_q}_0
                 \alpha(t)dt+\varepsilon(x_p-x_q)\ .
\label{AppA3}
\end{eqnarray}
In our approach, the kernels $K_N^{(\beta)}$, i.e. the complete upper
left entries of the $2\times2$ matrices in Eqs.~(\ref{AppA2}) and
(\ref{AppA2a}), are the fundamental quantities, rather than Mehta's
kernels.  Therefore the operators in the off diagonal elements should
also be defined as acting on $K_N^{(\beta)}$. This is accomplished by
the following definitions
\begin{eqnarray}
DK_N^{(\beta)}(x_p,x_q)&=&\frac{1}{2}\left(\frac{d}{dx_p}K_N(x_q,x_p)-
                \left(x_p\leftrightarrow x_q\right)\right)\nonumber\\
IK_N^{(\beta)}(x_p,x_q)&=&\frac{1}{2}\left(
            \int dt \varepsilon(x_p-t)K_N^{(\beta)}(t,x_q)-
            \left(x_p\leftrightarrow x_q\right)\right)\nonumber\\
JK_N^{(\beta)}(x_p,x_q)&=&IK_N(x_p,x_q)+\varepsilon(x_p-x_q)\ .
\label{AppA4}
\end{eqnarray}
With these definitions our expressions for the correlation functions,
Eqs.~(\ref{eq2.6}) and (\ref{eq2.7}), are identical with
Eqs.~(\ref{AppA2}) and (\ref{AppA2a}). This is easily verified by using
that $DS_N(x_p,x_q)$ and $IS_N(x_p,x_q)$ are antisymmetric in their
arguments \cite{Mehta}. We just remark that the simplicity of the
definitions~(\ref{AppA4}) is another strong hint that the functions
$K_N^{(\beta)}(x_p,x_q)$ rather than $S(x_p,x_q)$ are the fundamental
quantities.

\section{Real part contributions to the correlation functions}
\label{app2}

The correlation functions $R_k(x_1,\ldots,x_k)$ in classical RMT are,
according to Eq.~(\ref{eq2.0}), averages involving only the imaginary
parts of the Green functions. Including the real parts, one has the
more general correlation functions
\begin{equation}
\widehat{R}_k(x_1,\ldots,x_k) = \frac{1}{\pi^k} \int P_N(H)
\prod_{p=1}^k
                          \tr\frac{1}{H-x_p} d[H] \ .
\label{eq5.1}
\end{equation}
We use the notation of Ref.~\cite{Gu3}, cf.~Eq.~(\ref{eq3.4}). As for
the definition~(\ref{eq2.0}), we ignore contributions proportional to
$\delta(x_p-x_q)$. In the case of the GUE, it has been shown in
Ref.~\cite{Gu1} that the functions~(\ref{eq5.1}) also have a
determinant structure. We conjecture that the quaternion determinant
structure carries over to the GOE and the GSE cases, too. The
corresponding kernels $\widehat{K}_N^{(\beta)}(x_q,x_p)$ are
generalizations of the kernels $K_N^{(\beta)}(x_q,x_p)$. We expect that
they are given by
\begin{equation}
\widehat{K}_N^{(\beta)}(x_q,x_p) =
             \frac{1}{\gamma\pi}
             \frac{\exp\left(\gamma
                       \left(x_p^2-x_q^2\right)/2\right)}
                           {x_p-x_q}\nonumber\\
           \left(\frac{\beta^2}{2\gamma^4}
           \int \exp\left(-\frac{\beta}{2|\gamma|}\trg\sigma^2\right)
                      \detg^{-\beta N/2|\gamma|}(\sigma-x^-) d[\sigma]
                 -1 \right) \ ,
\label{eq5.2}
\end{equation}
such that Eq.~(\ref{eq3.3a}) results when taking the imaginary part. In
the GUE case $\beta=2$, formulae~(\ref{eq5.2}) is a immediate
consequence of Ref.~\cite{Gu1}. For the GOE and GSE cases $\beta=1$ and
$\beta=4$, formulae~(\ref{eq5.2}) states a conjecture.  The kernel
$\widehat{K}_N^{(2)}(x_q,x_p)$ follows from $K_N^{(2)}(x_q,x_p)$ by
simply replacing one of the oscillator wave functions $\varphi_n(z)$
with $\widehat{\varphi}_n(z)$. The latter function combines the two
independent solutions of the oscillator wave equation, i.e.~the
function $\varphi_n(z)$ and its Cauchy or Stiltjes transform. To the
best of our knowledge, the relevance of those second solutions in an
RMT context was first observed in Ref.~\cite{Gu2}. Again, we conjecture
that these features also carry over to the kernels
$\widehat{K}_N^{(\beta)}(x_q,x_p)$ and $K_N^{(\beta)}(x_q,x_p)$ for
$\beta=1$ and $\beta=4$.

\section{Calculation of some integration constants}
\label{app1}

Considering Eq.~(\ref{rec}) at $x_p = 0$, we obtain the recursion
formula
\begin{equation}
  c_N^{(1)} - \left(\frac{N}{2} + 1\right) c_{N+2}^{(1)} = \frac{4 \pi
  (-1)^{N+1}}{(N+1)!} \left( H_{N+1}(0) + N\,b_N - \frac{1}{2} \,
  b_{N+2} \right) \ ,
\label{intConstRec}
\end{equation}
where
\begin{equation}
b_N = \int_{-\infty}^{\infty} \varepsilon(t) \exp(-t^2/2) H_N(t) dt \ .
\end{equation}
The right hand side of Eq.~(\ref{intConstRec}) turns out to be zero for
all $N$. This is easily seen for even $N$. For odd $N$, one has to
employ Eq.~(\ref{part1}) and to integrate by parts. The recursion
formula obtained in this way is equivalent to a result given by
Mehta~\cite{Mehta}.  The remaining task is to find $c_0^{(1)}$ and
$c_1^{(1)}$ as starting values for an induction.

We employ the implicit definition of $\omega_1^{(1)}(x_p)$ according to
Eqs.~(\ref{Omega}) and~(\ref{fac}).  The difficulty is due to the
singularities. For $N=0$, it suffices to use
\begin{equation}
\frac{1}{s_{11}^{-}s_{21}^{-}} = \frac{1}{s_{11}^{-}-s_{21}^{-}}
\left(\frac{1}{s_{21}^{-}}-\frac{1}{s_{11}^{-}}\right) \ .
\end{equation}
A straightforward calculation and comparison with Eq.~(\ref{GOEparts})
gives
\begin{equation}
c_0^{(1)} = 0 .
\end{equation}
For $N=1$, the singular terms involve fractional exponents and the
steps needed are more complicated. One can employ an integral
representation of the $\Gamma$ function, valid for arbitrary $k$. It
yields
\begin{equation}
  \frac{1}{(s_{p1}^{-})^k} = \frac{i^k}{\Gamma(k)}
  \int_{0}^{\infty} dt t^{k-1} \exp(-its_{p1}^-) \ ,
\end{equation}
which moves the singularities into the exponent and decouples them from
the power $k$. Moreover, we introduce new integration variables
\begin{eqnarray}
t_1 = \frac{T+\tau}{2} \quad {\rm and} \quad t_2 = \frac{T-\tau}{2}
\quad {\rm with} \quad \tau = T \cos\vartheta \ .
\end{eqnarray}
All this leads to
\begin{eqnarray}
\frac{1}{\left(s_{11}^{-}s_{21}^{-}\right)^k} &=&
 \left(\frac{i^k}{\Gamma(k)}\right)^{2}  \int_{0}^{\infty} dt_{1}
  t_{1}^{k-1} \exp(-it_{1}s_{11}^-)
     \int_{0}^{\infty} dt_2
  t_{2}^{k-1} \exp(-it_{2}s_{21}^-) \nonumber \\
  &=& \left(\frac{i^k}{\Gamma(k)}\right)^{2} \frac{1}{2^{2k-1}}
      \int_{0}^{\infty} dT  T^{2k-1}
      \exp \left( -\frac{iT}{2}\left(s_{11}^{-} +
            s_{21}^{-}\right)\right)\nonumber \\
  & & \qquad
      \int_{0}^{\pi} d\vartheta \sin^{2k-1}\vartheta
      \exp\left(-\frac{iT\left(s_{11}-s_{21}\right)}{2}
                       \cos\vartheta\right)
                                      \nonumber \\
 &=& \frac{(-1)^k\sqrt{\pi}}{2^{(2k-1)/2}\Gamma(k)}
     \int_{0}^{\infty} dT  T^{2k-1}
     \exp \left( -\frac{iT}{2}\left(s_{11}^{-} +
              s_{21}^{-}\right)\right) \nonumber \\
 & & \qquad\qquad\qquad \frac{J_{(2k-1)/2}(T(s_{11}-s_{21})/2)}
     {(T(s_{11}-s_{21})/2)^{(2k-1)/2}} \ .
\label{b1}
\end{eqnarray}
Here, we did the angular integral using the representation
\begin{equation}
  \int _{0}^{\pi} \exp\left(iz\cos\vartheta\right)
     \sin^{d-2}\vartheta d\vartheta =
  2^{(d-2)/2} \sqrt{\pi} \Gamma \left(\frac{d-1}{2} \right)
  \frac{J_{(d-2)/2}(z)}{z^{(d-2)/2}} \ ,
\end{equation}
for the Bessel function in $d$ dimensions. We now insert Eq.~(\ref{b1})
into the implicit definition of $\omega_1^{(1)}(x_p)$ according to
Eqs.~(\ref{Omega}) and~(\ref{fac}). We also rotate the eigenvalues
\begin{eqnarray}
u = \frac{s_{11}+s_{21}}{2} \quad {\rm and} \quad v =
\frac{s_{11}-s_{21}}{2}
\end{eqnarray}
and find, at $x_p=0$,
\begin{eqnarray}
  \omega_1^{(1)}(0) &=&
  \im \int_{-\infty}^{+\infty} ds_{11}
      \int_{-\infty}^{+\infty} ds_{21}
      \frac{|s_{11} - s_{21}|}{\left(s_{11}^-s_{21}^-\right)^{3/2}}
  \exp\left(-\frac{1}{2}\left(s_{11}^{2}+s_{21}^{2}\right)\right)
                 \nonumber\\
  &=& \im \frac{i^32\sqrt{\pi}}{\Gamma(3/2)}
  \int_{-\infty} ^{+\infty} du\int_{-\infty} ^{+\infty} dv
              \int_{0}^{\infty} dT T \varepsilon(v)
                \exp\left(-u^2 - v^2\right)
               \exp(-iTu^-) J_1(Tv) \nonumber \\
  &=& -8 \pi + 4 \sqrt{2} \pi
\end{eqnarray}
with the step function $\varepsilon(v)$ defined in Eq.~(\ref{eq2.5}).
{}From Eq.~(\ref{GOEparts}), we also have
\begin{equation}
  \omega_1^{(1)}(0) = -8 \pi \int_{0}^{\infty}
           \exp\left(-\half t^2\right) t dt - c_1^{(1)}
  = -8\pi - c_1^{(1)} \ .
\end{equation}
Hence, combining the last two formulae, we obtain
\begin{equation}
c_1^{(1)} = -4 \sqrt{2} \pi
\end{equation}
for the case $N=1$.

Thus, we can now use the recursion~(\ref{intConstRec}) and finally
arrive at the result~(\ref{intconst}).

\end{document}